\documentclass[twocolumn]{aa} 

\pdfoutput=1
\usepackage{graphicx}
\usepackage{txfonts}
\usepackage{color}
\usepackage{amsmath}
\usepackage{amssymb}
\usepackage{rotating}
\usepackage{verbatim}
\usepackage{subfig}
\usepackage{booktabs}
\usepackage{longtable}
\usepackage[switch]{lineno}
\begin{document}

   \title{Venus Express radio occultation observed by PRIDE}


   \author{T. M. Bocanegra-Baham{\'o}n
          \inst{1,2,3}
          \and
          G. Molera\,Calv{\'e}s
          \inst{1,4}
          \and 
          L.I. Gurvits
          \inst{1,2}
	      \and
          G. Cim\`{o}
          \inst{1,5}
          \and 
          D. Dirkx
          \inst{2} 
          \and
          D.A. Duev
          \inst{6}
          \and
          S.V. Pogrebenko
          \inst{1}
          \and 
          P. Rosenblatt
          \inst{7}
          \and 
          S. Limaye
          \inst{8}
          \and
          L. Cui
          \inst{9}
          \and 
          P. Li
          \inst{9}
          \and 
          T. Kondo
          \inst{10,3}
          \and 
          M. Sekido
          \inst{10}
          \and 
          A.G. Mikhailov
           \inst{11}
          \and
          M.A. Kharinov
           \inst{11}
          \and
          A.V. Ipatov
           \inst{11}
          \and 
          W. Wang
          \inst{3}
          \and
          W. Zheng 
          \inst{3}
          \and 
          M. Ma
          \inst{3}
          \and 
           J.E.J. Lovell
          \inst{4}
          \and 
          J.N. McCallum
          \inst{4}
          }

   \institute{Joint Institute for VLBI ERIC, P.O. Box 2, 7990 AA Dwingeloo, The Netherlands.\\
              \email{bocanegra@jive.eu}\\
         \and
             Department of Astrodynamics and Space Missions, Delft University of Technology, 2629 HS Delft, The Netherlands.\\
         \and
             Shanghai Astronomical Observatory, 80 Nandan Road, Shanghai 200030, China. \\
 	 \and 
	     University of Tasmania, Private Bag 37, Hobart, Tasmania, 7001, Australia\\
           \and 
     Netherlands Institute for Radio Astronomy, P.O. Box 2, 7990 AA Dwingeloo, The Netherlands. \\       
	     \and
             California Institute of Technology, 1200 E California Blvd, Pasadena, CA 91125, USA.  \\
             \and 
             ACRI-ST, 260 route du Pin Montard, F06904 Sophia-Antipolis Cedex, France. \\
             \and
             Space Science and Engineering Center, University of Wisconsin, Madison, WI, USA.\\
             \and
             Xinjiang Astronomical Observatory, CAS, 150 Science 1-Street, Urumqi, Xinjiang 830011, China. \\
             \and
             NICT Kashima Space Technology Center, 893-1, Kashima, Ibaraki, 314-8501, Japan. \\
             \and 
             Institute of Applied Astronomy of Russian Academy of Sciences, 191187 St. Petersburg, Russia. }


 
  \abstract
   {Radio occultation is a technique used to study planetary atmospheres by means of the refraction and absorption of a spacecraft carrier signal through the atmosphere of the celestial body of interest, as detected from a ground station on Earth. This technique is usually employed by the deep space tracking and communication facilities (\emph{e.g.,} NASA's Deep Space Network (DSN), ESA's Estrack). }
  {We want to characterize the capabilities of the Planetary Radio Interferometry and Doppler Experiment (PRIDE) technique for radio occultation experiments, using radio telescopes equipped with Very Long Baseline Interferometry (VLBI) instrumentation.}
    {We conducted a test with ESA's Venus Express (VEX), to evaluate the performance of the PRIDE technique for this particular application. We explain in detail the data processing pipeline of radio occultation experiments with PRIDE, based on the collection of so-called open-loop Doppler data with VLBI stations, and perform an error propagation analysis of the technique.} 
   {With the VEX test case and the corresponding error analysis, we have demonstrated that the PRIDE setup and processing pipeline is suited for radio occultation experiments of planetary bodies. The noise budget of the open-loop Doppler data collected with PRIDE indicated that the uncertainties in the derived density and temperature profiles remain within the range of uncertainties reported in previous Venus' studies. Open-loop Doppler data can probe deeper layers of thick atmospheres, such as that of Venus, when compared to closed-loop Doppler data. Furthermore, PRIDE through the VLBI networks around the world, provides a wide coverage and range of large antenna dishes, that can be used for this type of experiments.}
   {}


   \maketitle
%

\section{Introduction}

The Planetary Radio Interferometry and Doppler Experiment (PRIDE) is a technique based on the adaptation of the traditional far-field very long baseline interferometry (VLBI) astrometric technique applied to near-field targets - spacecraft inside the solar system - with the main objective of providing precise estimates of the spacecraft state vectors. This is achieved by performing precise Doppler tracking of the spacecraft carrier signal and near-field VLBI observations in phase-referencing mode \citep{Duev2012,Duev2016,Bocanegra2017}. PRIDE is suitable for various applications in planetary and space science, such as determination of planetary ephemerides \citep{Dirkx2017}, characterization of the interplanetary plasma \citep{Molera2014} and detection of interplanetary coronal mass ejection (ICME) \citep{Molera2017}. In this work we present another application: the characterization of planetary atmospheres and/or ionospheres by means of radio occultation observations.

The implementation and application of radio occultation experiments to planetary science has been widely discussed in the literature \citep[\emph{e.g.}][]{Phinney1968,Fjeldbo1968,Fjeldbo1971,Jenkins1994,Yakovlev2002,Patzold2009,Tellmann2009,Withers2014}. A planetary radio occultation experiment involves a `central' body (a planet or a natural satellite), the atmosphere (or ionosphere) of which is to be studied, and two radio elements: a transmitter onboard a spacecraft orbiting (or performing a flyby about) the central body and, one or multiple ground stations on Earth. At certain geometries, the spacecraft is occulted by the central body with respect to the line of sight of the receiving ground station. As the spacecraft gets gradually occulted by the central body, the spacecraft carrier signal cuts through successively deeper layers of the planet's atmosphere, experiencing changes in its frequency and amplitude before being completely blocked by the body. This phase is known as ingress. Then the same phenomena is observed as the signal emerges from behind the body. This phase is known as egress. The refraction that the signal undergoes due to the presence of the atmosphere can be determined from the frequency shift observed throughout the occultation event. In addition to this, accurate estimates of the spacecraft state vector are needed to obtain the refractivity as a function of the radius from the occultation geometry. In the case of PRIDE, the receiving element is not a single station but a network of Earth-based radio telescopes. PRIDE provides multiple single-dish Doppler observables that are utilized to derive the residual frequencies of the spacecraft carrier signal, and additionally interferometry observables that are used, along with the Doppler observables, as input to determine the spacecraft state vector. The final product of this experiment is the derivation of vertical density, pressure, and temperature profiles of the central body's atmosphere.

The purpose of this paper is to evaluate the performance of the PRIDE setup for radio occultation experiments of planetary atmospheres. We note that the radio occultation technique can be productive for studies of atmosphereless celestial bodies as well (\emph{e.g.}, the Moon, Saturn's rings), enabling the characterization of the shape of the occulting body. However, we do not discuss this application here. As a test case, we analyze several observations of ESA's Venus Express (VEX) spacecraft using multiple radio telescopes from the European VLBI Network (EVN), AuScope VLBI array (Australia) and NICT (Japan), during April 2012, May 2012 and March 2014. The VEX orbiter was launched in 2005.11.09, arriving at Venus in April 2006 and remained in orbit around the planet until end of 2014. The operational orbit of the spacecraft was a highly elliptical ($e=0.99$), quasi-polar orbit ($i=\sim$90$^{\circ}$) with a $\sim$24-hour orbital period and a pericenter altitude of 250\,km. The radio science operations and analysis were led by the Venus Express radio science team (VeRa) \citep{Hausler2007}. The spacecraft, equipped with an onboard ultra-stable oscillator (USO) as frequency reference, conducted all radio occultation experiments in a one-way mode, transmitting coherent dual-frequency carrier downlinks at X-band (8.4\,GHz) and S-band (2.3\,GHz) with the main 1.3\,m high gain antenna (HGA1). The nominal mission receiving ground station during radio occultation observations was the 35-m Estrack New Norcia (NNO) station located in Western Australia. 

Besides VEX, Venus' atmosphere  and ionosphere has been studied with ground- and space-based telescopes,  and multiple spacecraft using a variety of techniques \citep{Brace1991, Fox1997, Limaye2017}. Using the planetary radio occultation technique, Venus' atmospheric and ionospheric density profiles have been derived from observations of NASA's Mariner 5 \citep{Mariner1967,Kliore1967,Fjeldbo1971}, Mariner 10 \citep{Howard1974}, Venera 9 and 10 \citep{Kolosov1979,Vasilyev1980}, Venera 15 and 16 \citep{Yakovlev1991,Gubenko2008}, Pioneer Venus Orbiter (PVO) \citep{Kliore1979,Kliore1980,Kliore1982,Kliore1985, Kliore1991}, Magellan \citep{Steffes1994,Jenkins1994,Hinson1995}, ESA's VEX \citep{Patzold2004,Patzold2009,Tellmann2009,Tellmann2012, Peter2014, Girazian2015} and recently with JAXA's Akatsuki \citep{Imamura2017,Ando2017}.  
The composition of Venus' ionosphere is primarily O$_2^{+}$ and smaller amounts of CO$_2^+$, O$^+$ and other trace species \citep{Girazian2015}. Although radio occultation experiments cannot be used to determine the specific composition of the ionosphere, the radio waves are sensitive to the electron distribution and therefore density profiles of the ionospheric plasma can be derived down to $\sim$100\,km. Similarly, for Venus' neutral atmosphere density and temperature profiles can be derived assuming the composition of the atmosphere (96.5\% CO$_2$, 3.5\% N$_2$) \citep{Kliore1985,Seiff1985}. In the case of Venus' atmosphere, radio occultation data provide a vertical coverage of $\sim$40-100\,km in altitude. This technique is the only remote sensing method that can probe Venus' atmosphere at altitudes below $\sim$65\,km \citep{Limaye2017}. 


In this paper we present the results of VEX occultation observations obtained with radio telescopes equipped with VLBI data acquisition instrumentation. We explain the processing pipeline carried out with the PRIDE setup and present the corresponding error propagation analysis. Sections \ref{ssec:radocctheoryS} and \ref{ssec:radoccgeom} present the theoretical background of the radio occultation method and the approximations taken into account when formulating the observation model. Section \ref{ssec:relationToAtmos} shows the theoretical derivation of the atmospheric profiles from the carrier signal frequency residuals. Section \ref{sec:testcase} presents the overall description of the experiment and the results found. Section \ref{ssec:expSetup} describes the setup of the experiment, describing the observations used and the data processing pipeline. Section \ref{ssec:atmosProfiles} present the resulting Venus' atmospheric profiles using the PRIDE setup. Section \ref{ssec:errorprop} presents the error propagation analysis, through the processing pipeline, from the Doppler observables to the derived atmospheric properties. Section \ref{sec:conclusions} presents the conclusions of this technology demonstration.

\section{The radio occultation experiment}
\label{sec:radocctheory}

In a radio occultation experiment, the carrier signal of the spacecraft experiences refraction, absorption and scattering as it passes behind the visible limb of the planetary body due to its propagation through the planet's atmosphere on its way to the receiving ground station on Earth. The physical properties of the planetary atmosphere can be inferred by analyzing the changes in frequency and amplitude of the received carrier signal. 

\subsection{Theoretical background and approximations}
\label{ssec:radocctheoryS}

During the occultation event, the spacecraft carrier signal propagates through the atmosphere of the celestial body, experiencing a modulation of phase and a decrease in amplitude. The variation in phase is proportional to the real part of the medium's complex refractive index and the decrease in amplitude is proportional to the imaginary part of the medium's complex refractive index, also known as absorption factor \citep{Eshleman1973}.  These refractive and absorptive radio effects are caused by the presence of neutral gases and plasma in the atmosphere and ionosphere. In the study at hand, only the frequency changes in the carrier signal are analyzed. Therefore, the amplitude data from which absorptivity profiles are derived, are not treated in this paper.

As the spacecraft signal gets refracted crossing the different layers of the planet's atmosphere, the variation of the real part of complex refractive index (treated simply as the refractive index in the remainder of this paper) as a function of altitude can be determined, for a particular cross section of the atmosphere. Before establishing the relation between the frequency changes of the received carrier signal and the planet's atmospheric refractive index as a function of altitude, let us consider some approximations that simplify the models used to relate these quantities.

As for most planets, the variations in the electrical properties of Venus' atmosphere occur at scales much larger than the spacecraft signal wavelength \citep{Fjeldbo1971}. Hence, the radio wave can be treated as a light ray and its trajectory can be determined by geometric optics. As in previous radio occultation experiments of Venus, such as \citet{Fjeldbo1971,Jenkins1994,Tellmann2012}, we assume the planet's atmosphere can be modeled as a spherically symmetric medium, made of concentric shells, each of them with a constant refractive index. From geometric optics the propagation path of the signal can be described as a ray $\textbf{r}(s)$ parameterized by an arclength $s$ that satisfies the following differential equation \citep{Born1999},

\begin{equation}
\frac{d}{ds}\left( n \frac{d \textbf{r}}{ds} \right) = \nabla n 
\label{eq:drds}
\end{equation}
where $n(\textbf{r})$ is the refractive index, which for the case of spherical symmetry is a function of radial distance only. Hence, the ray trajectory, as it bends through the medium, can be traced given the medium refractive index as a function of the radius. This is known as the forward problem. However, we are interested in the inverse problem, where based on the ray path parameters, -the bending angle and the impact parameter-, the refractivity of profile of the atmosphere is retrieved.


\subsection{Observation model}
\label{ssec:radoccgeom}

In the radio domain the bending that the signal undergoes as it crosses the planet's ionosphere and neutral atmosphere cannot be measured directly. However, it can be retrieved from the frequency shift experienced by the received signal at the ground stations. In this section, we will first introduce the relation between the frequency shift and the ray parameters of the received signal, to then establish the relation between ray parameters and refractive index as a function of radius. 

\subsubsection{Observation geometry}

\begin{figure}[!htp]
    \centering
    \includegraphics[scale=0.1]{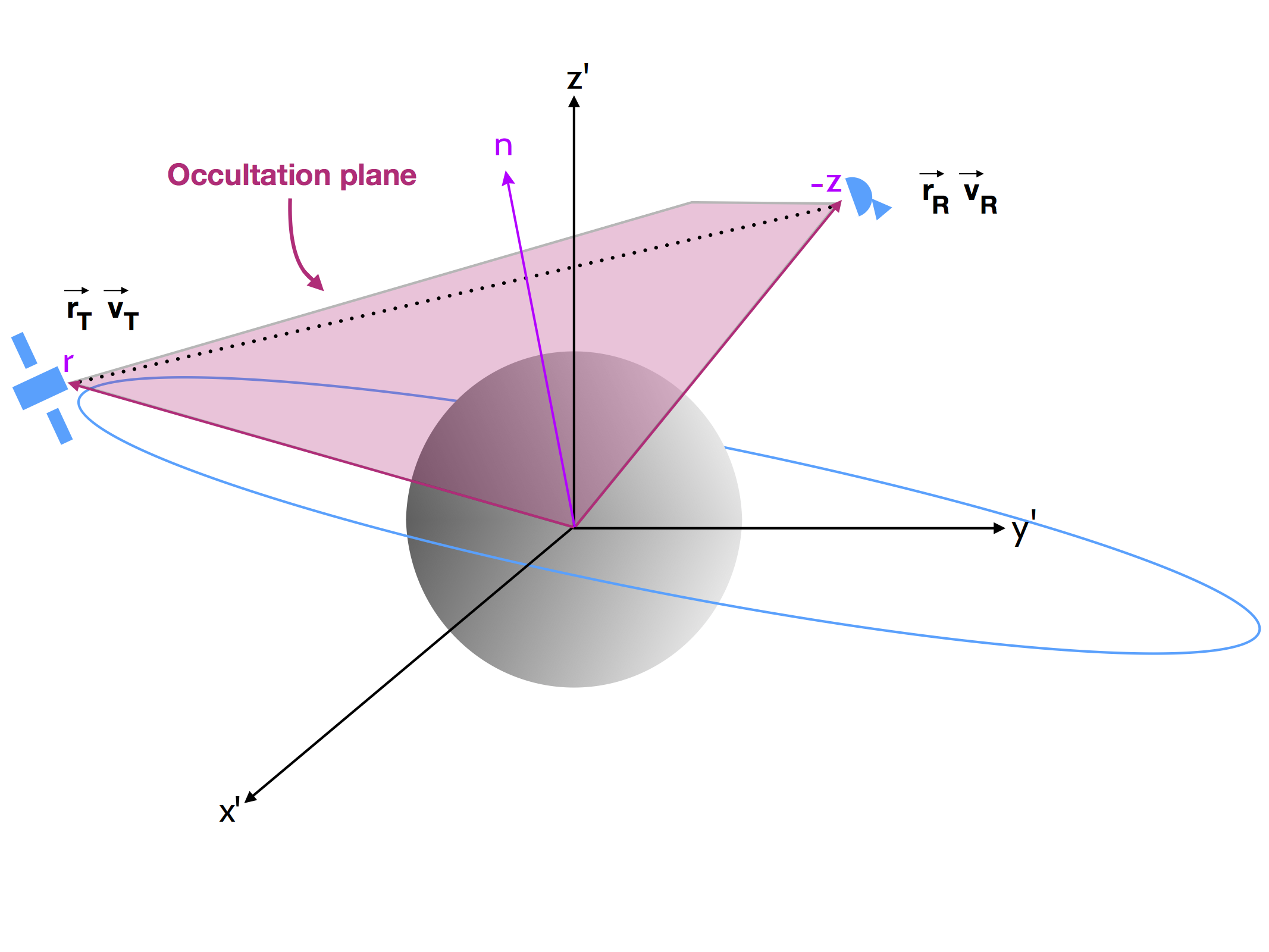}
    \caption[Sketch of the occultation plane.]{Sketch of the occultation plane. The occultation plane is defined by the center of mass of the target planet, the position of tracking station and the position of the spacecraft, each with respect to the target planet.}
    \label{fig:occultationPlane}
\end{figure}

The geometry of the occultation is determined by the occultation plane (\emph{i.e.}, the plane defined by the center of the target planet, the position of tracking station and the position of the spacecraft, each with respect to the target planet) as shown in Figure \ref{fig:occultationPlane}. The reference frame used has its origin at the center of mass of the target planet, with the negative $z$-axis pointing to the position of the tracking station at reception time, the $n$-axis parallel to the normal of the occultation plane and the $r$-axis perpendicular to both the $z$- and $n$- axis. In this scenario, the target body is assumed to have a spherically symmetric and stationary atmosphere and the ray path refraction occurs on the occultation plane, reducing it to a two-dimensional problem as shown in Figure \ref{fig:occultationGeometry}. The bending of the refracted ray path is then parameterized with respect to the free-space ray path by means of the angles $\delta_r$ and $\beta_r$. The angle $\delta_r$ is defined between the position vector of the spacecraft at transmission time with respect to the tracking station at reception time and the ray path asymptote in the direction the radio wave is received at the tracking station at reception time. The angle $\beta_r$ is defined between the position vector of the tracking station at reception time with respect to the spacecraft at transmission time and the ray path asymptote in the direction the radio wave is transmitted by the spacecraft at transmission time (see Figure \ref{fig:occultationGeometry}).

This description of the occultation geometry is based on \citet{Fjeldbo1971}, where the approach that will be discussed in this section was first introduced. Multiple authors have expanded on this approach \citep[\emph{e.g.,}][]{Lipa1979, Jenkins1994, Withers2014} 
 including the relativistic corrections into the analysis. 

\subsubsection{Derivation of the ray path parameters from the carrier signal frequency residuals}
\label{ssec:derivdeltaf}

To isolate the perturbation that the spacecraft signal experiences when propagating through the media along the radio path, we evaluate the difference between the detected carrier frequency and the prediction of the received frequency at the ground station, assuming for the latter that the signal is propagating through free-space (including geometrical effects, such as relative position and motion between the spacecraft and ground station, Earth rotation and relativistic corrections).  If perturbations due to the signal propagation through the Earth's atmosphere, ionosphere and interplanetary medium are accounted for, then the remaining observed perturbation is solely due to the atmosphere and ionosphere of the planet of interest.

As shown by \citet[Eq. 266]{Kopeikin1999}, the frequency received at a tracking station on Earth $f_R$ at a reception time $t_R$ is given by,

\begin{equation}
f_{R} = f_T \frac{1 - {\bf k}_R \cdot {\bf v}_R/c }{1 - {\bf k}_T \cdot {\bf v}_T/c} R(\mathbf{\mathbf{v}_{R},\mathbf{v_{T},t_{R},t_{T}}})\label{eq:fRfT}
\end{equation}
where $c$ is the free-space velocity of light, $f_T$ is the spacecraft transmission frequency at the transmission time $t_T$, ${\bf v}_R$  and ${\bf v}_T$ are the barycentric velocity vectors of the receiving station at $t_R$ and of the spacecraft at $t_T$, respectively, ${\bf k}_R$ and ${\bf k}_T$ are the unit tangent vectors in the direction along which the radio wave propagates at $t_T$ and $t_R$, respectively, and the term $R$ gives the special and general relativistic corrections:

\begin{equation}
R ( v_R, v_T, t_R, t_T)= \left[ \frac{ 1 - (v_T/c)^2}{ 1 - (v_R/c)^2} \right]^{1/2} \left[ \frac{a(t_T)}{ a(t_R)} \right]^{1/2} \frac{b(t_R)}{b(t_T)}
\end{equation}
where the terms $a$ and $b$ are derived as explained in Section 2.2 of \citet{Bocanegra2017}. All position and velocity vectors are expressed in the solar system barycentric frame.

The frequency residuals are subsequently found by evaluating the difference between detected frequency $f_{R,detected}$ and the predicted frequency in free-space $f_{R,free-space}$ received at the ground station at $t_R$, as follows:

\begin{equation}
\Delta f = f_{R, detected} - f_{R, free-space} 
\label{eq:deltaf}
\end{equation}
assuming that $f_{R,detected}$ has been corrected for the effects of propagation through interplanetary plasma and the Earth's atmosphere and ionosphere.  

\begin{figure*}[!htp]
    \centering
    \includegraphics[width=0.8\textwidth]{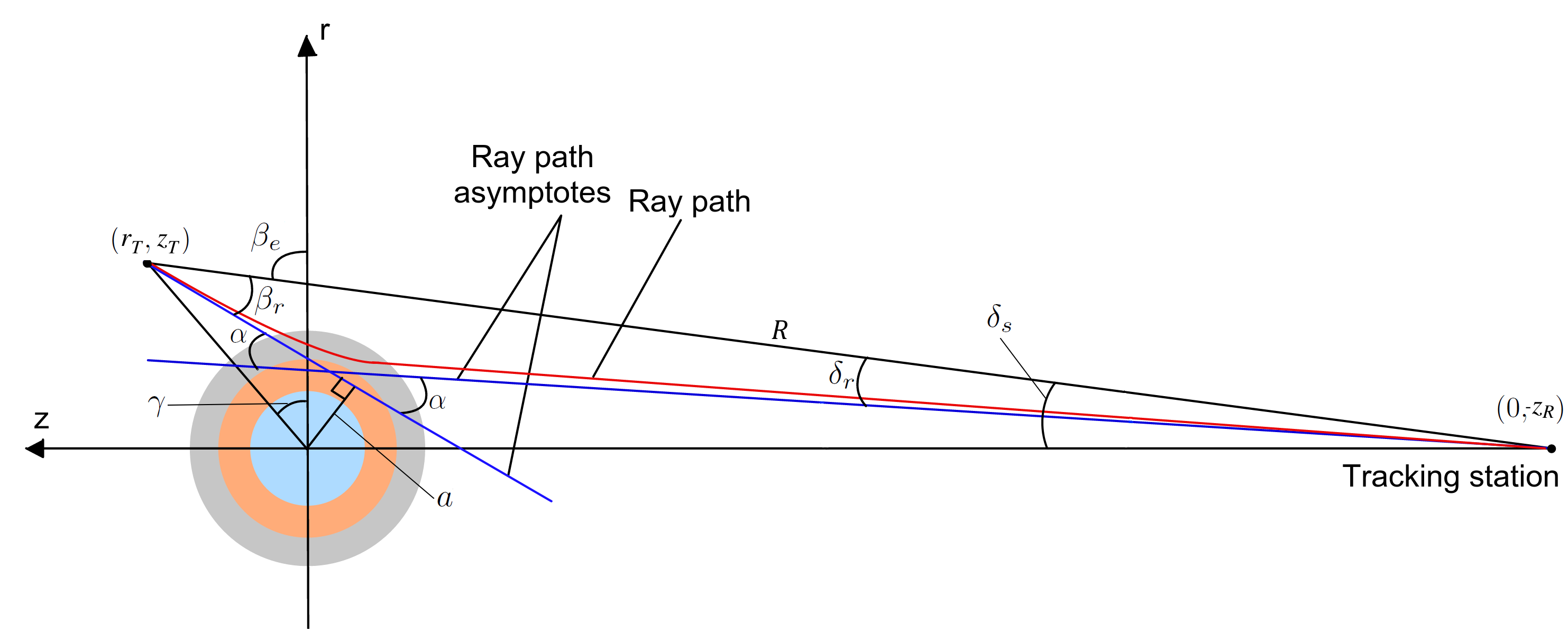}
    \caption[View of the geometry of the radio occultation, depicted on the occultation plane.]{Geometry of the radio occultation depicted on the occultation plane. In this case, the figure shows the ray path as it gets refracted by the planet's ionosphere (in gray), where it gets bent by an angle $\alpha$ from its original path.}
    \label{fig:occultationGeometry} 
\end{figure*}

In the case of free-space, the direction along which the radio signal propagates is the same at $t_R$ and $t_T$, hence ${\bf k}_R = {\bf k}_T$ in Eq \ref{eq:fRfT}, and $\delta_r = 0$ and $\beta_r=0$. When the signal gets refracted by the atmosphere of the target planet, ${\bf k}_R$ and ${\bf k}_T$ become the two ray path asymptotes shown in blue in Figure \ref{fig:occultationGeometry}. Hence, the ray path asymptotes for both cases can be defined as follows:

\begin{equation}
-{\bf k}_{R, free-space} = {\bf \hat{r}} \sin{\delta_s} + {\bf \hat{z}} \cos{\delta_s}
\label{eq:krfreespace}
\end{equation}
\begin{equation}
-{\bf k}_{T, free-space} = {\bf \hat{r}}\cos{\beta_e } + {\bf \hat{z}} \sin{\beta_e}
\end{equation}
\begin{equation}
-{\bf k}_{R, detected} = {\bf \hat{r}}\sin{(\delta_s - \delta_r)} + {\bf \hat{z}} \cos{(\delta_s - \delta_r)} 
\end{equation}
\begin{equation}
-{\bf k}_{T, detected} = {\bf \hat{r}} \cos{(\beta_e - \beta_r)} + {\bf \hat{z}} \sin{(\beta_e - \beta_r)} 
\label{eq:ktdetected}
\end{equation}
%


Hence, a relation between the frequency residuals and the angles that parameterize the bending of the ray path can be established by replacing Eq. (\ref{eq:fRfT}) and Eqs. (\ref{eq:krfreespace}) to (\ref{eq:ktdetected}) into Eq. (\ref{eq:deltaf}),

\begin{align}
\Delta f & = R f_T \left ( \frac{ c + v_{r,R} \sin{(\delta_s - \delta_r)} + v_{z,R} \cos{(\delta_s - \delta_r)} }{ c + v_{r,T}\cos{(\beta_e - \beta_r)} + v_{z,T} \sin{(\beta_e - \beta_r)} } \right. \nonumber \\
 & \left. - \frac{ c + v_{r,R} \sin{\delta_s} + v_{z,R}\cos{\delta_s} }{ c + v_{r,T} \cos{\beta_e } + v_{z,T} \sin{\beta_e} } \right ) 
\label{eq:deltafpro}
\end{align}
where $v_{r,R}$ is the $r$-component of $v_R$ and $v_{z,R}$ is the $z$-component of $v_R$ at reception time $t_R$, $v_{r,T}$ is the $r$-component of $v_T$  and $v_{z,T}$ is the $z$-component of $v_T$ at transmission time $t_T$. The angles $\beta_e$ and $\delta_s$ are defined between the position vector of the spacecraft at $t_T$ with respect to the ground station at $t_R$, and the $r$-axis and $z$-axis, respectively ($ \beta_e + \delta_s = \pi/2$).


As a consequence of Eq. (\ref{eq:drds}), the distance $a$ from the planet's center of mass to the tangent at any point along the ray path is a constant $a$,

\begin{equation}
a = |z_R|\sin{(\delta_s - \delta_r)}=(r_T^2 + z_T^2)^{1/2}\sin{(\beta_e - \gamma - \beta_r)}
\label{eq:a}
\end{equation}
where $a$ is called the ray impact parameter.
Eqs. (\ref{eq:deltafpro}) and (\ref{eq:a}) can be solved simultaneously for the two ray path angles $\delta_r$ and $\beta_r$ - assuming that the state vectors of the spacecraft and tracking station are known - using the numerical technique introduced by \citet{Fjeldbo1971}. Once the values of $\delta_r$ and $\beta_r$ are determined, the total bending angle of refraction $\alpha$ of the ray path is obtained by adding them (see Figure \ref{fig:occultationGeometry}). 

\begin{equation}
\alpha = \delta_r + \beta_r
\label{eq:alpha}
\end{equation}
Applying this procedure, for every $\Delta f$ obtained at each sampled time step (Eq. \ref{eq:deltaf}) the corresponding ray path parameters, bending angle $\alpha$ and impact parameter $a$, are derived. We follow the sign convention adopted by \citet{Ahmad1998} and \citet{Withers2010} where positive bending is considered to be toward the center of the planet.

\subsection{Relation to atmospheric properties}
\label{ssec:relationToAtmos}

In this section we will establish the relations that link the ray path parameters to the physical properties of the layer of the atmosphere the radio signal is sounding. Following \citet{Born1999}, assuming a radially symmetric atmosphere represented by $K$ concentric spherical layers each with a constant refractive index $n$, the bending angle $\alpha$ is related to $n$ through an Abel transform:

\begin{equation}
\alpha(a_k) = -2a_k \int\limits^{r = \infty}_{r = r_k} \frac{d \ln( n(r))}{ dr} \frac{dr}{\sqrt{ ( n(r)r)^2 - a_k^2}}
\label{eq:abel}
\end{equation}
where $r$ is the radius from the center of the planet to the ray and the subscripts represent the $k$-th layer, the deepest layer each ray path reaches, which corresponds to a specific reception time at the receiver.

As explained in \citet{Fjeldbo1971}, Eq. (\ref{eq:abel}) can be inverted to have an expression for the refractive index $n$ in terms of $\alpha$ and $a$:

\begin{equation}
\ln{ n_k(a_k) } = \frac{1}{\pi}\int\limits_{a'=a_k}^{a'=\infty} \frac{\alpha(a')da'}{\sqrt{a'^2 - a^2_k}}
\label{eq:abelinv}
\end{equation}

\begin{equation}
n(r_{0k}) = \exp{\left[ \frac{1}{\pi} \int^{\alpha'=0}_{\alpha'=\alpha(a_k)} \ln\left[ \frac{a(\alpha')}{a_k} + \left[ \left( \frac{a(\alpha')}{a_k} \right)^2 - 1 \right]^{1/2} \right] d\alpha' \right]}.
\label{eq:muR0k}
\end{equation}
Using the total refraction bending angles found as explained in Section \ref{ssec:radoccgeom}, the refractive index for the $k$-th layer can be determined by performing the integration over all the layers the ray has crossed:

\begin{equation}
n_k(a_k) = \exp{ \left\{ \frac{1}{\pi}\int\limits_{a_1}^{a_0}\frac{\tilde{\alpha}_1da'}{\sqrt{a'^2 - a^2_k}} + ... +  \frac{1}{\pi}\int\limits_{a_k}^{a_{k-1}}\frac{\tilde{\alpha}_k da'}{\sqrt{a'^2 - a^2_k}} \right \} }
\label{eq:solveAbel}
\end{equation}
where $\tilde{a}_i$ is the average value of the bending angles $\alpha_i$ in a layer:

\begin{equation}
\tilde{\alpha}_i = \frac{ \alpha_i(a_i) + \alpha_{i-1}(a_{i-1})}{2} \mbox{ with } i=1...k.
\end{equation}
A ray with impact parameter $a$ will cross the symmetric atmosphere, down to a layer $k$ of radius $r_{0k}$. This minimum radius is found via the Bouguer's rule, which is the Snell law of refraction for spherical geometries. As explained in \citet{Born1999} and \citet{Kursinski1997}, $r_{0k}$ is related to $ a_k$ by:

\begin{equation}
r_{0k} = \frac{a_k}{n_k}
\end{equation}
The refractivity as a function of the radius depends on the local state of the atmosphere. The total refractivity of the atmosphere $\mu$ is given by the sum of the components due to the neutral atmosphere and ionosphere. For each layer the total refractivity is given by \citep{Eshleman1973}:

\begin{equation}
\mu_k = (n_k - 1) = \mu_{n,k} + \mu_{e,k}
\label{eq:mu_k}
\end{equation}
where $\mu_n$ is the refractivity of the neutral atmosphere:

\begin{equation}
\mu_n = \kappa N_n
\label{eq:mun}
\end{equation}
where $\kappa$ is the mean refractive volume and $N_n$ is the neutral number density, and $\mu_e$ is the refractivity of the ionosphere:

\begin{equation}
\mu_e = -\frac{N_ee^2}{8\pi^2m_e\epsilon_0f^2}
\label{eq:mue}
\end{equation}
where $e$ is the elementary charge, $m_e$ is the electron mass, $\epsilon_0$ is the permittivity of free-space, $f$ is the radio link frequency and $N_e$ is the electron density.

In the ionosphere, the electron density is high and the neutral densities can be several orders of magnitude lower, therefore in the ionosphere $\mu_e$ is dominant and $\mu_n$ is negligible. On the other hand, at lower altitudes, the situation is the opposite: the neutral densities are high and electron densities are low. Hence in practice, if the value of $\mu$ is negative, $\mu_e$ is assumed to be equal to $\mu$ and if $\mu$ is positive, then $\mu_n$ is assumed to be equal to $\mu$.

Assuming hydrostatic equilibrium, the vertical structure of the neutral atmosphere can be derived from the neutral density profile $N_n(h)$ and the known constituents of the planetary atmosphere. The pressure in an ideal gas is related to the temperature $T(h)$ and number density of the gas by:

\begin{equation}
p(h) = k N_n(h)T(h)
\label{eq:idealgas}
\end{equation}
where $k$ is the Boltzmann's constant. Using Eq. (\ref{eq:idealgas}) and the equation for hydrostatic equilibrium the temperature profile can be found from the following formula \citep{Fjeldbo1968}:

\begin{equation}
T(h) = T(h_0)\frac{N(h_0)}{N(h)} + \frac{\bar{m}}{kN(h)}\int\limits_{h}^{h_0}g(h')N(h')dh' 
\label{eq:tempN}
\end{equation}
where $\bar{m}$ is the mean molecular mass, $g(h)$ is the gravitational acceleration and $h_0$ is an altitude chosen to be the top of the atmosphere for which the corresponding temperature $T(h_0)$, taken from the planet's reference atmospheric model, is assigned as boundary condition. From Eq.(\ref{eq:tempN}) it can be seen that the sensitivity of $T(h)$ to the upper boundary condition $T(h_{0})$ rapidly decreases due to the $N(h_0)/N(h)$ factor. 

%
%

\section{PRIDE as an instrument for radio occultation studies: a test case with Venus Express}
\label{sec:testcase}

The PRIDE technique uses precise Doppler tracking of the spacecraft carrier signal at several Earth-based radio telescopes and subsequently performs VLBI-style correlation of these signals in the so-called phase referencing mode \citep{Duev2012}. In this way, PRIDE provides open-loop Doppler observables, derived from the detected instantaneous frequency of the spacecraft signal \citep{Bocanegra2017}, and VLBI observables, derived from the group and phase delay of the spacecraft signal \citep{Duev2012}, that can be used as input for spacecraft orbit determination and ephemeris generation. During a radio occultation experiment, the Doppler observables are utilized to derive the residual frequencies of the spacecraft carrier signal. These are subsequently used to derive atmospheric density profiles, as explained in Section \ref{sec:radocctheory}. Both, the Doppler and VLBI observables, can be used for orbit determination, allowing the accurate estimation of the spacecraft state vectors during the occultation event. In this paper, we will focus on the error propagation in the frequency residuals obtained with the open-loop Doppler observables derived with PRIDE, and will use the VEX navigation post-fit orbits derived by the European Space Operations Center (ESOC)\footnote{ftp://ssols01.esac.esa.int/pub/data/ESOC/VEX/}, which do not include PRIDE observables, to calculate the spacecraft state vectors in the occultation plane.

\subsection{Observations and experimental setup}
\label{ssec:expSetup}

\begin{figure}[!htp]
    \centering
    \includegraphics[scale=0.25]{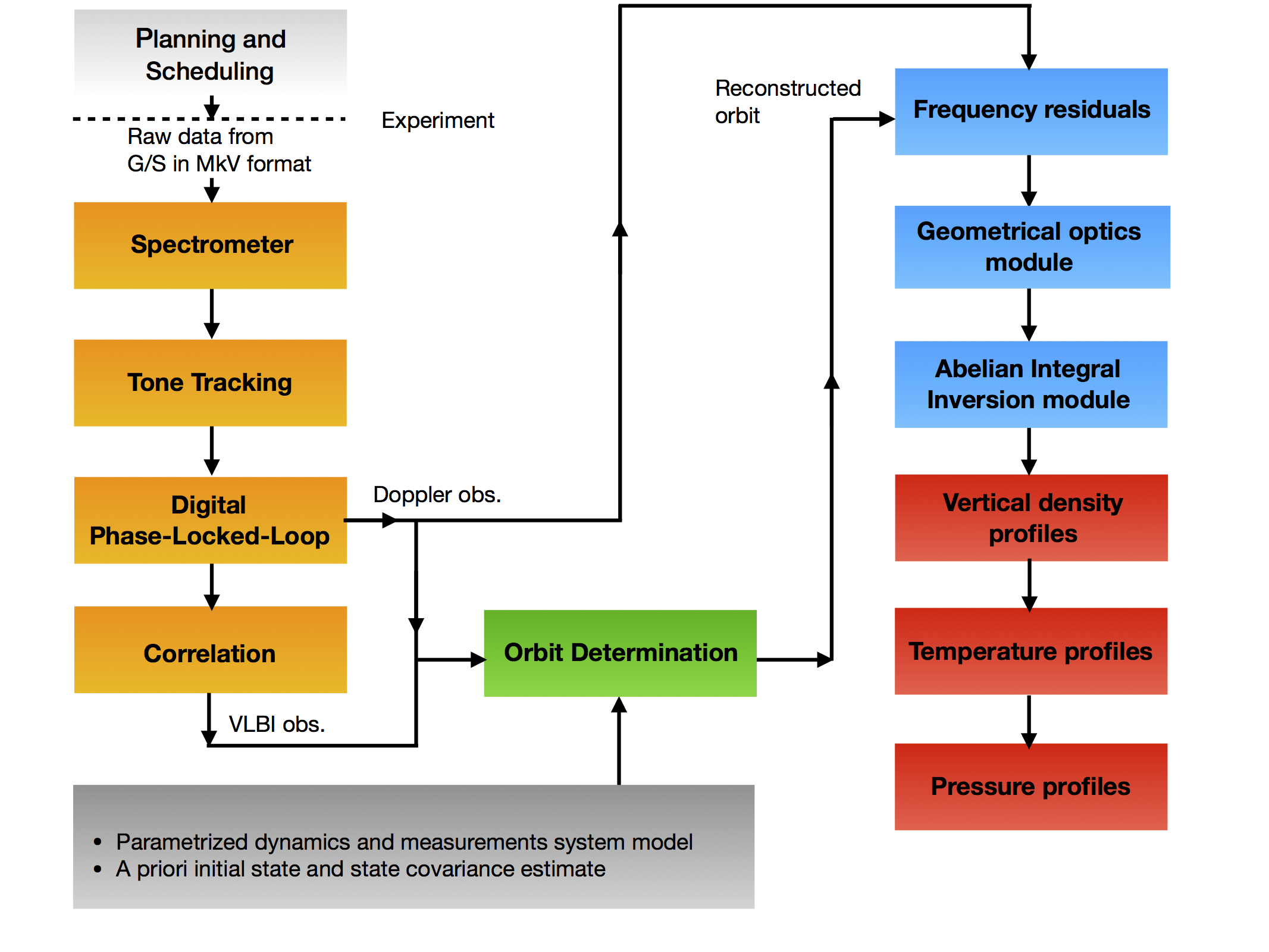}
    \caption{Radio occultation processing pipeline. The first part of the software, which comprises the \texttt{SWSpec}, \texttt{SCtracker}, \texttt{PLL} and \texttt{SFXC} correlation software, (in yellow blocks in Figure \ref{fig:pipeline}) is used for every standard PRIDE experiment. This part of the processing pipeline gives open-loop Doppler and VLBI observables as output. The second part of the software was developed for the sole purpose of processing radio occultation experiments. It consists of three main modules: the frequency residuals derivation module, the geometrical optics module and the Abelian integral inversion module. The output of this second part are vertical density, temperature and pressure profiles of the target's atmosphere. }
    \label{fig:pipeline}
\end{figure}

The current data processing pipeline of the PRIDE radio occultation experiment is represented in Figure \ref{fig:pipeline}. The first part of the software comprises three software packages developed to process spacecraft signals (the software spectrometer \texttt{SWSpec}, the narrowband signal tracker \texttt{SCtracker} and the digital phase-locked loop \texttt{PLL})\citep{Molera2014}\footnote{https://bitbucket.org/spacevlbi/} and the software correlator \texttt{SFXC} \citep{Keimpema2015} which is able to perform VLBI correlation of radio signals emitted by natural radio sources and spacecraft. These four parts of the processing pipeline are used for every standard PRIDE experiment (yellow blocks in Figure \ref{fig:pipeline}). The output at this point are open-loop Doppler and VLBI observables. The methodology behind the derivation of these observables using the aforementioned software packages is explained by \citet{Duev2012, Molera2014, Bocanegra2017}. The second part of the software was developed for the sole purpose of processing radio occultation experiments. It consists of three main modules: the frequency residuals derivation module, the geometrical optics module and the Abelian integral inversion module. From these three modules the vertical density profiles, and subsequently, temperature and pressure profiles of the target's atmosphere can be derived. 
The frequency residuals module uses the output of the \texttt{PLL} which are the time averaged carrier tone frequencies detected at each telescope. To produce the frequency residuals the predictions of the received carrier signal frequency at each telescope is computed as described in \citep{Bocanegra2017}, and then the frequency residuals are corrected with a baseline fit to account for the uncertainties in the orbit used to derive the frequency predictions (Section \ref{ssec:errorprop}).  In the geometrical optics module the state vectors
retrieved from VEX navigation post-fit orbit are transformed into a coordinate system defined by the occultation plane as explained in Section \ref{sec:radocctheory}, and using the frequency residuals found in the previous step, the bending angle and impact parameter at each sample step is found using Eqs. (\ref{eq:deltafpro}) to (\ref{eq:alpha}) using the procedure described in Section \ref{sec:radocctheory}. The refractivity profile is derived in the Abelian integral inversion module, where the integral transform that relates the bending angle with the refractive index (Eq. \ref{eq:abel}) is solved by modeling the planet's atmosphere as K concentric spherical layers of constant refractivity and applying Eq. (\ref{eq:solveAbel}), as explained in Section \ref{sec:radocctheory}. The number of layers and their thickness is defined by the integration time step of the averaged carrier frequency detections. In the atmospheric model, each sample is assumed to correspond to a ray passing tangentially through the middle of the $n$-th layer, as described in \citet{Fjeldbo1971} (Appendix B).

\begin{figure*}[!htp]
    \centering
    \includegraphics[width=0.8\textwidth]{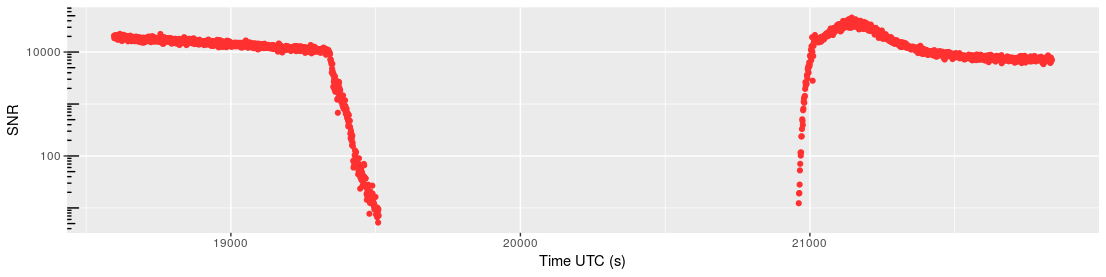}
    \caption[Example of the S/N of a signal detection during a radio occultation.]{Example of the S/N of a signal detection during a radio occultation. This is the VEX signal detection obtained with Bd in the session of 2012.04.29. At around 19400\,s the S/N starts rapidly decreasing marking the beginning of the occultation ingress, which lasts for $\sim$3 minutes before there is loss-of-signal (LOS).  At around 20950\,s there is acquisition-of-signal (AOS) marking the beginning of the occultation egress which lasts for $\sim$1.5 minutes. The peak of the detected S/N after egress corresponds to the closest approach of VEX to the center of mass of Venus. A higher S/N is typically observed during VEX's radio science observation phase (scheduled around the pericenter passage) because the telemetry is off during this phase. During the tone tracking part of the processing, in order to provide an appropriate polynomial fit to the low S/N part of the detection, the ingress and egress scan are split in two. For this particular example, for the ingress scan at 19400\,s and for the egress scan at 21080\,s.}
    \label{fig:LOS}
\end{figure*}

To provide a noise budget of PRIDE for radio occultation experiments we used five observation sessions in X-band (8.4\,GHz) with Earth-based telescopes between 2012.04.27 and 2012.05.01, and one in 2014.03.23, when VEX was occulted by Venus. Table \ref{tab:names} shows a summary of the telescopes involved in the observations and Table \ref{tab:summary} shows a summary of the observations and participating telescopes per day. The observations were conducted in three scans\footnote{A scan is the time slot in which the antenna is pointing to a specific target (\emph{i.e}, a spacecraft or a natural radio source)}: in the first scan the antennas pointing to the spacecraft for 19 minutes up to ingress where there is loss of signal, in the second the antennas pointing to the calibrator source for 4 minutes, and in the third pointing back to the spacecraft starting right before egress for 29 minutes. The exception being the 2014.03.23 session where the ingress and egress were detected in one single scan (no calibrator source was observed in between, because there was no loss of signal throughout the occultation). Figure \ref{fig:LOS} shows the signal-to-noise ratio (S/N) of the two spacecraft scans, ingress and egress, as recorded by the 32-m Badary telescope on 2012.04.30. Usually, for orbit determination purposes, the part of the scan where the S/N of the detections starts dropping due to the signal refraction in the planet's atmosphere is discarded. However, this is precisely the part of the scan that is of interest for radio occultation experiments. For the case of VEX, during radio occultation sessions the spacecraft was required to perform a slew manoeuvre to compensate for the severe ray bending due to Venus' thick neutral atmosphere \citep{Hausler2006}. 

As explained in \citet{Bocanegra2017}, when processing the observations a polynomial fit is used to model the moving phase of the spacecraft carrier tone frequencies along the time-integrated spectra per scan. In order to provide an appropriate polynomial fit to the low S/N part of the detection, after running \texttt{SWSpec} the scan is split in two parts: right before the signal starts being refracted in the case of the ingress, and right after it stops being refracted in the case of the egress (Figure \ref{fig:LOS}). The initial phase polynomial fit and the subsequent steps with \texttt{SCtracker} and \texttt{DPLL} are conducted as if they were two separate scans. 

\begin{table}[!htbp]
\caption{Summary of the radio telescopes involved in the observations.}
\centering
\footnotesize 
\begin{tabular}{@{}l c c c @{} } 
\toprule
\multicolumn{ 1}{c}{Observatory} & \multicolumn{ 1}{c}{Country}       & \multicolumn{ 2}{c}{Telescope} \\ \cmidrule{ 3-4 } 
\multicolumn{ 1}{l}{} & \multicolumn{ 1}{l}{} & Code & Diameter (m) \\
\midrule
Sheshan (Shanghai) & China & Sh & 25 \\ \midrule
Nanshan (Urumqi) & China & Ur & 25  \\ \midrule
Tianma & China & T6 & 65 \\ \midrule
Badary & Russia & Bd & 32 \\ \midrule
Katherine & Australia & Ke & 12 \\ \midrule
Kashima & Japan & Ks & 34 \\
\bottomrule
\end{tabular}
\label{tab:names}
\end{table}

\begin{table*}[!htbp]
\caption[Summary of observations.]{Summary of observations.}
\centering
\begin{tabular}{@{} c c c c @{} } 
\toprule
Station time & Telescopes & Solar elongation & Distance to S/C \\
(UTC)&  Code      & (deg)            & (AU)            \\   
\midrule
2012-04-27 05:10 - 06:05 & Ur,Sh & 41  & 0.47 \\ \midrule
2012-04-29 05:10 - 06:05 & Bd,Ke & 40  & 0.46 \\ \midrule
2012-04-30 05:10 - 06:05 & Bd    & 40  & 0.45 \\ \midrule
2012-05-01 05:10 - 06:05 & Bd,Ks & 39  & 0.44 \\ \midrule
2014-03-23 02:50 - 03:16 & T6    & 46  & 0.68 \\
\bottomrule
\end{tabular}
\label{tab:summary}
\end{table*}

\begin{figure*}[!htp]
    \centering
    \subfloat[]
    {\includegraphics[height= 0.3\textwidth, width= 0.4\textwidth]{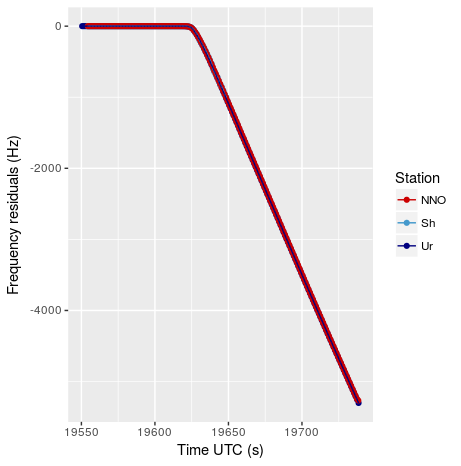}
     \label{fig:freqRes}}
     \subfloat[]
    {\includegraphics[height= 0.3\textwidth, width= 0.35\textwidth]{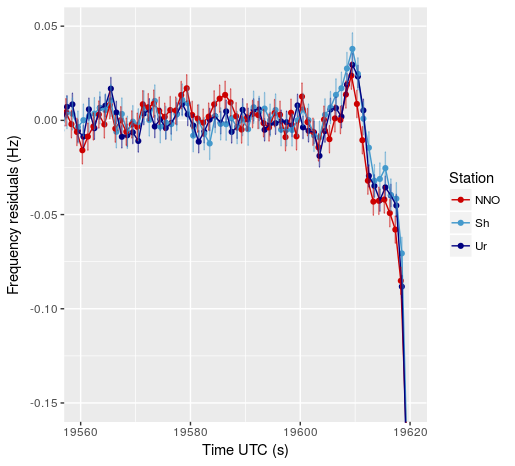}
    \label{fig:freqResZoom} } \\
    \subfloat[]
    {\includegraphics[height= 0.3\textwidth,width= 0.35\textwidth]{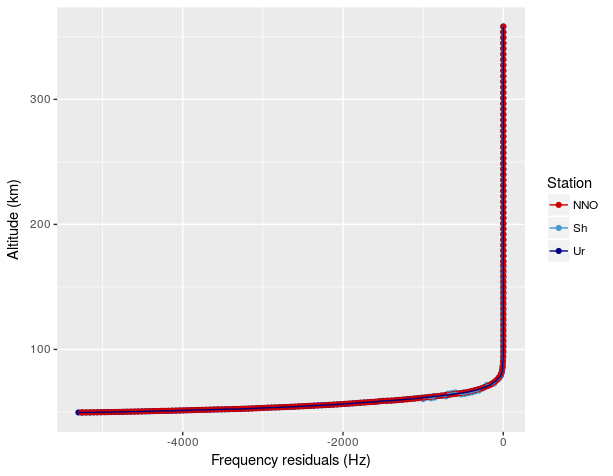}
     \label{fig:freqResImpParam}}
     \subfloat[]
    {\includegraphics[height= 0.3\textwidth,width= 0.4\textwidth]{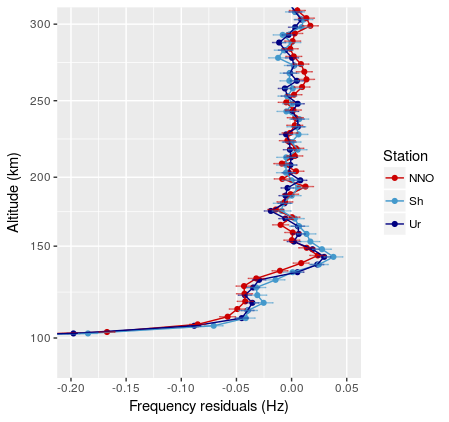}
    \label{fig:freqResZoomImpParam} } \\
    \subfloat[]
    {\includegraphics[height= 0.3\textwidth,width= 0.4\textwidth]{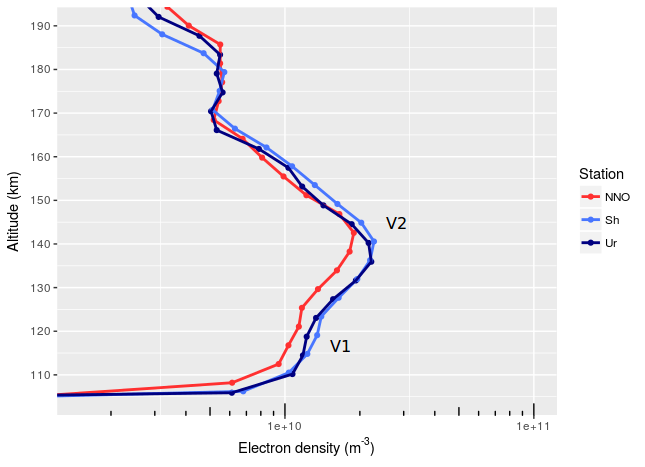}
    \label{fig:ionoElectronDensity} }
     \caption[Frequency residuals retrieved from open-loop data from Ur and Sh  compared to the residuals from closed-loop data from NNO, during the session of 2012.04.27.]{Frequency residuals retrieved from open-loop data from Ur and Sh  compared to the residuals from closed-loop data from NNO, during occultation ingress in the session of 2012.04.27. Panel (a) shows the frequency residuals of the detected signal up to LOS. Panel (b) zooms in the frequency residuals showing when the signal starts getting refracted by Venus' ionosphere at around 19605\,s. Panel (c) and (d) show the corresponding altitude probed versus the frequency residuals of the data shown in panel (a) and (b), respectively. Panel (e) shows the corresponding electron density profile, where the secondary V1 layer and main V2 layer of Venus' ionosphere are identified.}
\end{figure*}

\subsection{Derived atmospheric profiles}
\label{ssec:atmosProfiles}

Figure \ref{fig:freqRes} shows an example of the frequency residuals found for Ur and Sh during ingress for the 2012.04.27 session in comparison with those of NNO, as provided by ESA's planetary science archive (PSA) \footnote{\url{ftp://psa.esac.esa.int/pub/mirror/VENUS-EXPRESS/VRA/}}. From $\sim$19605\,s, the signal starts being refracted by the ionosphere, as shown in Figure \ref{fig:freqResZoom}, and from $\sim$19615\,s, where the frequency residuals start rapidly decreasing, marks the immersion into the neutral atmosphere. The largest frequency residuals (derived as explained in Section \ref{ssec:derivdeltaf}) observed in the depth of Venus' neutral atmosphere at X-band usually reach levels of $\sim$5\,kHz before loss of signal, as shown in Figure \ref{fig:freqRes}. Figure \ref{fig:ionoElectronDensity} shows the electron density profile derived from the frequency residuals shown in Figures \ref{fig:freqResZoom} and \ref{fig:freqResZoomImpParam}. In Figure \ref{fig:ionoElectronDensity} the ionospheric base can be identified at 100 km, along with two layers. The main layer has density $\sim$ 2$\times10 ^{10}$\,m$^{-3}$ and altitude $\sim$140\,km and the secondary layer has density 
$\sim$ 1$\times10 ^{10}$\,m$^{-3}$ and altitude $\sim$125\,km. This nightside (SZA = 142 deg) profile is consistent with other observations of the deep nightside ionosphere (\emph{e.g.}, \citet{Kliore1979,Patzold2009}). 

\begin{figure*}[!htp]
  \centering    
  \includegraphics[height= 0.35\textwidth,width= 1.0\textwidth]  
  {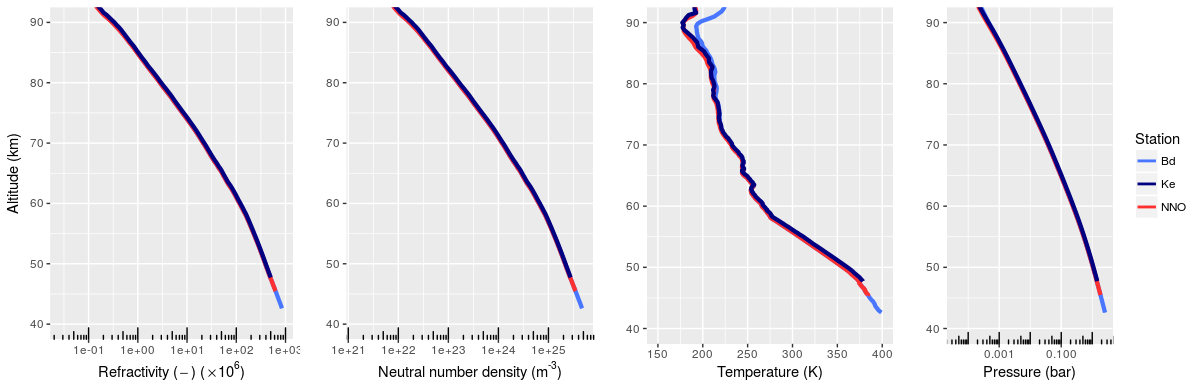} 
  \caption[Refractivity, neutral number density, temperature and pressure profiles of the 2012.04.29 session.]{Refractivity, neutral number density, temperature and pressure profiles of the 2012.04.29 session. The profiles corresponding to the 32-m Bd and 12-m Ke were derived from the open loop Doppler data obtained with the PRIDE setup, and the profiles of the 35-m NNO were derived using the frequency residuals obtained from ESA's PSA, corresponding to closed loop Doppler tracking data. Despite the fact that both NNO and Bd have similar antenna dish sizes, with Bd the spacecraft signal is detected down to a lower altitude, due to the fact that the Doppler data obtained with Bd is open loop, while for NNO is closed loop.}
  \label{fig:allPlots120429} 
\end{figure*}

\begin{figure}[!htp]
    \centering
  \includegraphics[width= 0.48\textwidth]{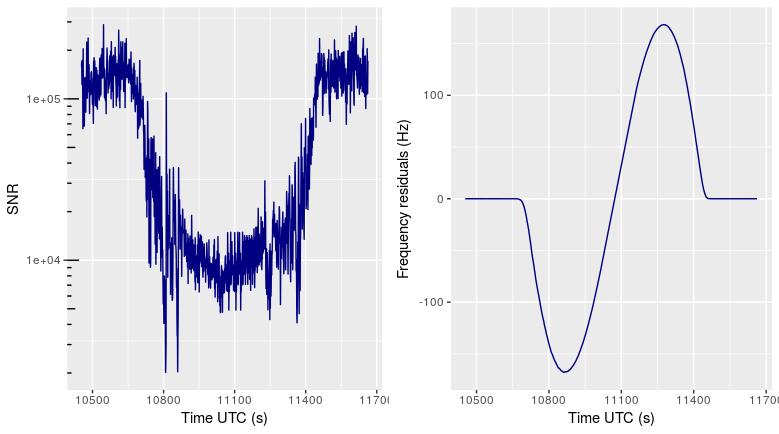}
    \caption[Detection of the carrier signal
of the spacecraft throughout a complete
occultation, using the open-loop Doppler data retrieved by the 65-m T6 during the session of 2014.03.23.]{Detection of the carrier signal
of the spacecraft throughout a complete
occultation, using the open-loop Doppler data retrieved by the 65-m T6 during the session of 2014.03.23. The left panel shows the S/N throughout the detection, showing that there is no LOS during the whole occultation. The right panel shows the frequency residuals, which show from 10650 to 11100\,s the ingress followed by the egress from 11100 to 11450\,s. This means that T6 was able to detect the signal while the planetary disk was completely occulting VEX.}
    \label{fig:T6}
\end{figure}

Figure \ref{fig:allPlots120429} shows an example of the resulting ingress profiles of Venus' neutral atmosphere from the Doppler frequency residuals corresponding to the session of 2012.04.29. Figure \ref{fig:allPlots120429} shows the refractivity, neutral number density, temperature and pressure profiles derived with Eqs. (\ref{eq:mu_k}), (\ref{eq:mun}), (\ref{eq:tempN}) and (\ref{eq:idealgas}), respectively. Despite the fact that both NNO and Bd have similar antenna dish sizes, with Bd the spacecraft signal is detected down to a lower altitude. During this observation the elevation of Bd was between 50-58\,deg while for NNO it was 20-25\,deg. Several noise contributions at the antenna rapidly increase, such as the atmospheric and spillover noise, at low elevation angles, which result in lower S/N detections. Besides this, the profiles corresponding to Bd and Ke were derived from the open loop Doppler data obtained with the PRIDE setup, while the profiles of NNO were derived using the frequency residuals obtained from ESA's PSA, corresponding to closed loop Doppler tracking data. The advantage of using open loop Doppler data for radio occultation resides in the ability of locking the signal digitally during the post-processing. This allows the estimation of the frequency of the carrier tone at the deeper layers of the atmosphere. This is not the case with closed-loop data, since once the system goes out of lock the signal is lost. Figure \ref{fig:T6} shows the frequency detections obtained by the 65-m Tianma during the session of 2014.03.23, where the carrier signal of the spacecraft is detected throughout a complete occultation, including the time slot where the planetary disk is completely occulting the spacecraft.
This is possible because of the extremely strong refraction the signal undergoes while crossing Venus' neutral atmosphere. The fact that there is no loss-of-signal (LOS) in this particular session highlights the advantage of using both large antenna dishes and open loop data processing for radio occultation experiments. On one hand, the large antenna dishes have low thermal noise allowing higher S/N detections throughout the occultation, and on the other hand, the open-loop processing allows the detection of the carrier signal through thick media. It is worth to mention, however, that for the session 2014.03.23 the closed-loop Doppler data of NNO is not currently publicly available, therefore Tianma's results could not be compared with those of NNO.


\begin{table*}[!htbp]
\caption[Summary of radio occultation sessions.]{Summary of location and depth of the radio occultation profiles obtained during the different sessions.}
\centering
\begin{tabular}{@{} c c c c c c @{} } 
\toprule
Date &  Station  & Mode    & Minimum altitude & Latitude & LST        \\
     &           &         & (km)            &  (deg)   & (hh:mm:ss) \\ 
\midrule
2012-04-27  & Ur & Ingress & 46.2           &   -20.2  & 01:48:19  \\
            &    & Egress  & 43.7           &    83.4  & 23:19:29  \\
            & Sh & Ingress & 45.8           &   -20.0  & 01:48:08  \\
            &    & Egress  & 44.7       	&    83.5  & 23:08:12  \\\midrule
2012-04-29  & Bd & Ingress & 42.7           &   -19.7  & 01:59:55  \\
            &    & Egress  & 44.0           &    83.6  & 23:17:13   \\
            & Ke & Ingress & 47.8           &   -23.6  & 01:57:49   \\
            &    & Egress  & 44.3           &    83.6  & 23:12:25   \\ \midrule
2012-04-30  & Bd & Ingress & 41.0           &   -19.3  & 02:05:49 \\
            &    & Egress  & 42.0           &    83.7  & 23:12:39   \\ \midrule
2012-05-01  & Bd & Ingress & 41.9       	&   -22.0  & 02:10:27 \\
			&    & Egress  & 43.1		    &   83.7   & 23:35:13 \\
            & Ks & Ingress & 45.1     	    &  -26.8   & 02:07:31 \\
            &    & Egress  & 47.2			&	83.8   & 23:13:37 \\ \midrule
2014-03-23  & T6 & Ingress & 54.9          	& 	33.3   &  00:02:07  \\
           & 	 & Egress  & 54.9		    &	33.3   & 00:02:07	\\
\bottomrule
\end{tabular}
\label{tab:summaryprofiles}
\end{table*}

\begin{figure*}[!htp]
    \centering
    \includegraphics[scale=0.7]{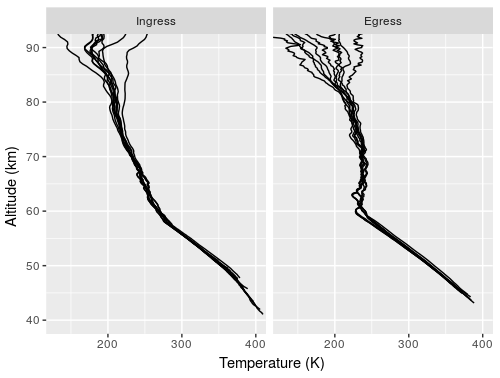}
    \caption[Compilation of the temperature profiles obtained with all participating station during the session of 2012.04.27 to 2012.05.01.]{Compilation of the temperature profiles obtained with all participating stations during the session of 2012.04.27 to 2012.05.01. The left panel shows the ingress temperature profile corresponding to latitudes around 20\,deg S. The right panel shows the egress temperature profiles corresponding to latitudes around 84\,deg N.}
    \label{fig:TAll}
\end{figure*}

Table \ref{tab:summaryprofiles} gives the location and depth of the radio occultation profile obtained during the different sessions. Figure \ref{fig:TAll} shows the neutral atmosphere temperature profiles obtained using Eq. (\ref{eq:tempN}) throughout the different sessions in 2012, displaying the ingress and egress profiles separately.  Besides the assumption of a spherically symmetric atmosphere, the radio occultation method as discussed in this paper, also assumes hydrostatic equilibrium and a known composition. The composition is assumed to be constant in a spherically homogeneous well-mixed atmosphere below the altitude of the homopause (<$\sim$125\,km). For the analysis in this paper, atmospheric composition of the neutral atmosphere is assumed to be 96.5\% CO$_2$, 3.5\% N$_2$ \citep{Kliore1985,Seiff1985}. In order to derive the temperature profiles, an initial guess for the boundary temperature ($T(h_0)$ in Eq. \ref{eq:tempN}) of 170\,K at $100$ km altitude was used. In the case of Venus, the boundary temperature is typically chosen to be between 170-220\,K at 100\,km altitude, with an uncertainty of 35\,K. These values are taken from the temperature profiles of the Venus International Reference Atmosphere (VIRA) model \citep{Seiff1985,Keating1985} and the empirical model of Venus' thermosphere (VTS3) \citep{Hedin1983}.

As shown in Figure \ref{fig:TAll}, the troposphere is probed down to altitudes of $\sim$41\,km. From this altitude to about $\sim$58\,km the temperature decreases as altitude increases. Performing a linear fit from 43 to 58 \,km, the mean lapse rate found for the egress profiles is 9.5\,K/km. For the ingress profiles, a fit was performed from 41 to 50\,km that resulted in a mean lapse rate of 6.4\,K/km and another fit was made from 50 to 58\,km that resulted in a mean lapse rate of 9.3\,K/km. From the VIRA model, the mean lapse rate between 41-58\,km from the surface is 9.8\,K/km \citep{Seiff1985}. 
At about $\sim$58\,km the linear trend experiences a sharp change, which is identified as the tropopause, where the upper boundary of the middle clouds is present \citep{Seiff1985,Tellmann2012}. From the altitudes 60 to 80\,km, where the middle atmosphere extends, there is a clear difference between the ingress and egress profiles. This is due to the difference in latitudes of the occultation profiles, which are $\sim$20$^{\circ}$S and $\sim$83$^{\circ}$N for the ingress and egress, respectively, for the sessions from 2012.04.27 to 2012.05.01. For the egress profiles, the region between $\sim$65 to $\sim$70\,km is approximately isothermal, and from 70\,km to 80\,km the temperature decreases at a much lower lapse rate than at the troposphere. For the ingress profile, this isothermal behavior only extends for a couple of kilometers from $\sim$62\,km. The difference in lapse rate between the ingress and egress profiles along the altitude range from 60 to 80\,km can be attributed to dynamical variations on local scales, eddy motions or gravity waves \citep{Hinson1995}. It is desired to start the integration of the temperature profiles (Eq. \ref{eq:tempN}) as high as possible. However, at high altitudes (above $\sim$100\,km) there is not enough neutral gas detected and the noise introduced by the measurements is large with respect to the estimated refractivity values. For this reason, the upper boundary is chosen to be at an altitude of $\sim$100\,km, where the standard deviation of the refractivity is $\sim1/10$th of the estimated values. As shown by Eq. \ref{eq:tempN}, the sensitivity of the derived temperature profiles to the upper boundary condition $T(h_0)$ rapidly decreases with altitude due to the factor $N(h_0)/N(h)$. Therefore, it was found that when using different upper boundary temperatures (170\,K, 200\,K and 220\,K) at 100\,km for the detections of a single station, the temperature profiles of that particular station converge at $\sim$90\,km. When using the same upper boundary for all stations participating in one observation (\emph{e.g.}, the temperature profile shown in Figure \ref{fig:allPlots120429}c for the 2012.04.29 session), the temperature profiles of the different stations converge at $\sim$80\,km. This is due to the effect of the noise fluctuations of the refractivity profile, where the standard deviation in refractivity drops from 10$\%$ of the estimated value at 100\,km to $\sim0.1\%$ at $\sim$80\,km.

\subsection{Error propagation analysis}
\label{ssec:errorprop}

In order to quantify the performance of the PRIDE technique for the purpose of radio occultation experiments we carry out an error propagation analysis. To this end, we begin by propagating the frequency residual uncertainties derived from the open-loop Doppler data of the VLBI stations, through the multiple steps of data processing pipeline (Figure \ref{fig:pipeline}) to derive the uncertainties in the atmospheric properties of Venus measured during the radio occultation observations.

The frequency residual uncertainty $\sigma_{\Delta f}$ in time is the uncertainty of the difference between the observed frequency (also known as the sky frequency) $f_{sky}$, and the predicted frequency $f_{pred}$, the latter derived as explained in \citet{Bocanegra2017}. The $\sigma_{\Delta f}$ is evaluated before the signal starts getting refracted by the planet's atmosphere (referred to in this paper as the frequency residuals in `vacuum' for the sake of simplicity), that is, in the first part of the two-part split ingress scan, or the second part of the split egress scan. In practice, the standard deviation is calculated after performing a baseline correction to the frequency residuals in vacuum. While the uncertainty of the sky frequency is random, the uncertainty of the predicted frequency is systematic, reflecting the errors of the estimated state vectors of the spacecraft and ground stations, and the errors in the ephemerides of Venus and the Earth, used to generate the Doppler predictions, and the errors in the estimation of the nominal spacecraft transmission frequency at transmission time. In the absence of these systematic errors, the frequency residuals in vacuum would be the remaining Doppler noise of the signal (due to the random uncertainties) with zero mean value, as shown in Figure \ref{fig:freqRes} up to 19600\,s. The presence of systematic errors in the frequency residuals in vacuum results in a non-zero mean constant trend. This effect is corrected by applying a baseline correction based in the algorithm described by \citet{Gan2006}. For this reason, the $\sigma_{\Delta f}$, after baseline correction, is assumed to be solely due to $\sigma_{f_{sky}}$. 

The uncertainty of the $f_{sky}$ derived with the procedure shown in \citet{Bocanegra2017} depends on the integration time used, the noise introduced by the instrumentation on the ground stations and onboard the spacecraft, and the noise introduced by the propagation of the signal, $\sigma_{prop}$, through the interplanetary medium, and, the Earth's ionosphere and troposphere. In the case of one-way Doppler the instrumental noise is given by the thermal noise, $\sigma_{th}$, introduced by the receiver at the ground stations and the limited power received at downlink, the noise introduced by the spacecraft USO, $\sigma_{USO}$, the noise introduced by frequency standard used at the ground stations, $\sigma_{FS}$, and the antenna mechanical noise, $\sigma_{mech}$. The modeled $\sigma_{\Delta f}$ is calculated by,

\begin{equation}
\sigma^2_{\Delta f} = (\sigma^2_{th} + \sigma^2_{USO}  + \sigma^2_{FS} +  \sigma^2_{mech} + \sigma^2_{prop})f^2
\label{eq:variances}
\end{equation}
where $f$ is the carrier tone frequency which is $\sim$8.4\,GHz at X-band. The Allan deviation related to the thermal noise of the ground station receivers is given by \citet{Barnes1971,Rutman1991} as,

\begin{equation}
\sigma_{th}(\tau) \approx \sqrt{3BS_{\phi}}/2\pi f_0 \tau
\label{eq:sigmatau}
\end{equation}
where $S_{\phi}$ is the one-sided phase noise spectral density of the received signal in a 1\,Hz bandwidth, $B$ is the carrier loop bandwidth which is 20\,Hz for this experiment, $f_0$ is the nominal frequency of the Doppler link and $\tau$ is the integration time, which in this case has the value of 1\,s. The relative noise power of the carrier tone is given by $S_{\phi}$, which is approximated by $1/(\mbox{S/N})$ where S/N is the signal-to-noise ratio of the signal evaluated in a 1\,Hz bandwidth. The thermal noise values estimated with Eq. (\ref{eq:sigmatau}) of the X-band Doppler detections for the different radio telescopes used during the study at hand are summarized in Table \ref{tab:thermalNoise}. The uncertainties found correspond to values ranging from 1.6\,mHz for the lowest thermal noise, found at the 65-m T6, to 10.4\,mHz for the highest thermal noise, found at the 34-m Ks (higher than the one found for the 12-m Ke). As reported by \citet{Hausler2007}, the USO that provided the reference frequency for the VEX transponder has an Allan deviation of $\sim3 \times 10^{-13}$ at integration times 1-100\,s, which corresponds to $ \sigma_{USO}f < 2.5$\,mHz at 1\,s integration time. Based on these values the modeled $\sigma_{\Delta f}$ for each station is calculated using Eq. (\ref{eq:variances}) and displayed in Table \ref{tab:thermalNoise}, along with the measured $\sigma_{\Delta f}$ of the signal detections at 1\,s integration time (derived by taking the mean standard deviation of the frequency residuals in vacuum). All VLBI stations are equipped with a hydrogen maser frequency standard that provide a stability of $<1 \times 10^{-13}$ at $\tau=1$\,s, corresponding to a $\sigma_{FS}f < 0.9$\,mHz. From the instrumental noises, the antenna mechanical noise has not been taken into account due to the lack of information for these structures in the time intervals relevant for the study at hand. 

Regarding the noise due to propagation effects, the noise induced by the Earth's ionosphere is calibrated using the total vertical electron content (vTEC) maps \citep{Hernandez2009} and the noise introduced by the Earth's troposphere is calibrated using the Vienna Mapping Functions VMF1 \citep{BOEHM06}. These two corrections are applied to the Doppler predictions before deriving the frequency residuals $\Delta f$. The noise introduced by the remaining errors after calibration with the vTEC and VMF1 maps are not quantified in the work at hand. Furthermore, the interplanetary plasma noise was not characterized in this study due to the fact that only X-band detections were obtained. However, an estimate of the Allan deviation of the interplanetary phase scintillation noise can be derived using the approach described in \citet{Molera2014}. Based on the analysis of multiple tracking campaigns of the VEX signal, in S- and X-band, at different solar elongations, a relation (Eq. 6 in \citet{Molera2014}) was formulated to estimate the expected phase variation of the received signal as a function of TEC,

\begin{equation}
\sigma_{\mathrm{expected}} = \frac{\mathrm{TEC}}{4000}\cdot \left( \frac{8.4 \mathrm{GHz}}{f_{\mathrm{obs}}} \right)\cdot \left( \frac{\tau_{\mathrm{nod}}}{300s} \right)^{\frac{m+1}{2}} \mathrm{ [rad],}
\label{eq:sigmaprop1}
\end{equation}
where TEC is Venus-to-Earth total electron content along the line-of-sight, $\tau_{\mathrm{nod}}$ is the nodding cycle of the observations, $f_{\mathrm{obs}}$ is the observing frequency and $m$ is the spectral index. Since the phase power spectrum $S_{\phi}$ is of the form $S_{\phi} = Af^{-m}$, where $m$ is the spectral index and $A$ is a constant, the scintillation contribution in the spectral power density can be characterized using a first order approximation on the logarithmic scale, as explained in \citet{Molera2014} (Eq. 5). Using this approximation the expected phase variation can be expressed as follows,

\begin{equation}
\sigma_{\mathrm{expected}} = \left [ \frac{ A f_c^{-m} \cdot f_c}{ m + 1} \right ]^{\frac{m+1}{2}}
\label{eq:sigmaprop2}
\end{equation}
where $f_c$ is the cut-off frequency of the spectral power density of the phase fluctuations (usually $\sim$0.2\,mHz).
Following \citet{Armstrong1979}, for $2<m<4$ the Allan variance of a phase spectrum of the form $S_{\phi} = Af^{-m}$ is \citep{Armstrong1979}, 

\begin{equation}
\sigma_y^2(\tau) = \frac{A \tau^{m}}{\pi^2 f_{0}^2 \tau^3} \int^{\infty}_0 \frac{\sin^4{(\pi z)}}{z^m}dz
\label{eq:sigmaprop3}
\end{equation}
Based on the results presented in \citet{Molera2014}, it is assumed
a Venus-to-Earth TEC along the line-of-sight of $10^2$ tecu at 40\,deg elongation (Fig. 6 of \citet{Molera2014}) and a spectral index of $2.4$, which is the average spectral index of more than hundred observing sessions. Using these values and Eqs. \ref{eq:sigmaprop1} to \ref{eq:sigmaprop3} the estimated Allan deviation is of $2.2 \times 10^{-13}$ at $\tau = 1$\,s, which corresponds to $\sigma_{prop}f= 1.8$\,mHz. 

\begin{figure}[!htp]
    \centering
    \includegraphics[scale=0.6]{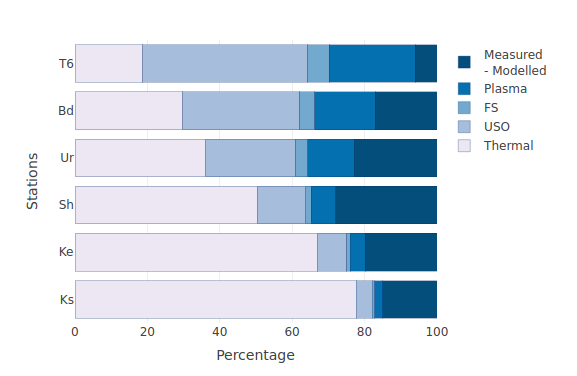}
    \caption[Percentage of different noise sources in the total measured frequency residuals.]{Percentage of the modeled noise sources in the total measured frequency residuals. In the case of T6 the predominant noise comes from the USO. For Bd, both the thermal noise and the USO noise represent each 30$\%$ of the total frequency residuals measured. Although Ur and Sh have the same antenna size the thermal noise of Sh is higher than that of Ur. The modeled thermal noise of Ke corresponds to almost 70\% of the budget, which in comparison to other stations is reasonable since its antenna diameter is only 12\,m and the station has an uncooled receiver. On the other hand, in the case of the 34-m Ks, the thermal noise corresponds to 80\% of the budget which given its size indicates an under-performance of the station during the observations it was involved in. In the overall budget, the noise introduced by the frequency standard is considered to be marginal with respect to the other sources of noise.}
    \label{fig:barchat}
\end{figure}

Figure \ref{fig:barchat} shows the percentages of the Allan variances of each modeled noise source with respect to the total measured frequency residuals variance for each station. The dark blue portion of the bar represents the difference between the measured and the modeled variances. In the case of T6 the predominant noise comes from the USO. For Bd, both the thermal noise and the USO noise represent each $\sim$30$\%$ of the total frequency residuals measured. This indicates that for one-way radio occultation experiments the noise introduced by the USO is larger than the noise introduced by the thermal noise of the ground stations, when using antennas with a diameter larger than $\sim$30\,m. This limiting noise source could be avoided by performing two-way tracking during occultations, however this would limit the occultation detections only to ingress passes since the transmitting signal would be out of lock during egress. 
The thermal noise of Sh is estimated to be higher than the thermal noise of Ur, which coincides with the larger frequency residuals obtained with Sh, despite the fact that both stations have the same antenna diameter.
The thermal noise of Ke corresponds to $\sim$70$\%$ of the noise budget. This is reasonable when compared to the thermal noise of the other stations, given the fact the Ke has a 12-m diameter. On the other hand, the modeled thermal noise for Ks corresponds to almost 80$\%$ of its noise budget, which is extremely high for a 34-m antenna. This was corroborated with the system temperature readings for the session of 2012.05.01 (also the only session where Ks participated), which were much higher than its reported nominal system temperature \citep{EVNSTATUS}. For this reason the data retrieved with Ks should be discarded. The modeled noise attributed to interplanetary plasma phase scintillation, estimated for elongation angles of $\sim$40\,degrees, can be corrected for when multifrequency observations are available \citep{Bertotti1993, Tortora2004}. For T6 this would correspond to a calibration of $\sim$20$\%$ of the Doppler noise. The noise introduced by the ground stations frequency standard in the overall noise budget is marginal compared to the other sources of noise. 

As shown in Table \ref{tab:thermalNoise} the modeled noise for the different telescopes is consistently lower than the measured noise. This can be attributed to unmodeled noise (\emph{e.g.}, antenna mechanical noise) and errors in the estimates of the propagation noise and the noise introduced by the USO. The estimate of the propagation noise given in Table \ref{tab:thermalNoise} includes only the plasma scintillation noise, but not the remaining errors after calibration of the Earth's ionospheric and tropospheric propagation effects. As for the USO, the Allan deviation given in Table \ref{tab:thermalNoise} represents the nominal frequency stability of the USO, hence, not the actual measurements during the observations.



Regarding the choice of integration time to process the radio occultation scans, there is a trade-off to be made between vertical resolution and frequency residual noise. The choice of a small integration time results in a better vertical resolution, but also results in larger frequency residuals noise. Furthermore, the vertical resolution of the profiles derived using geometrical optics are diffraction limited to the Fresnel zone diameter \citep{Hinson1999}, which is $\approx$2$\sqrt{\lambda D}$, where $\lambda$ is the signal wavelength, $D = R \cos(\beta_e - \gamma - \beta_r)$, and $R$, $\beta_e$, $\gamma$ and $\beta_r$ are defined as shown in Figure \ref{fig:occultationGeometry}. To be consistent with this inherited vertical resolution limit of the model the sample spacing should be kept as close to $\sqrt{\lambda D}$ as possible, which for this experiment should be larger than $\sim$470\,m. When processing the signal detections, it was noticed that the minimum integration time for which this condition would be satisfied was 0.1\,s down to $\sim$50\,km from the surface, since from this altitude on the bending angle of the ray path largely increases. Hence, the integration time was chosen to be 0.1\,s down to an altitude of 50\,km and then from this point down to lowest altitudes probed an integration time of 1.0\,s was used.

\begin{table*}[!htbp]
\caption[Noise budget of X-band Doppler detections of the VLBI stations during the radio occultation sessions.]{Noise budget of X-band Doppler detections of the VLBI stations during the radio occultation sessions.}
\centering
\footnotesize 
\begin{tabular}{@{}c c c c c c c c c c c c @{} } 
\toprule
\multicolumn{ 1}{c}{Station} & \multicolumn{ 2}{c}{Thermal} &
\multicolumn{ 2}{c}{USO} & \multicolumn{ 2}{c}{Frequency} &
\multicolumn{ 2}{c}{Plasma phase} &
\multicolumn{ 1}{c}{Modeled} & \multicolumn{ 1}{c}{Measured}\\
\multicolumn{ 1}{l}{} & \multicolumn{ 2}{c}{noise} &
 \multicolumn{ 2}{c}{ } &  \multicolumn{ 2}{c}{standard} &
  \multicolumn{ 2}{c}{scintillation} &
$\sigma_{\Delta f}$  & $\sigma_{\Delta f}$ \\ \cmidrule{ 2-9 } 
\multicolumn{ 1}{l}{} & $\sigma_{th}$ & $\sigma_{th}f$&
 $\sigma_{USO}$ & $\sigma_{USO}f$&
 $\sigma_{FS}$ & $\sigma_{FS}f$&
 $\sigma_{prop}$ & $\sigma_{prop}f$& 
   &   \\ 
\multicolumn{ 1}{l}{} &  & (mHz) &
  & (mHz) &   & (mHz) &   & (mHz) & (mHz) & (mHz) \\
\midrule
T6 &  $1.9 \times 10^{-13}$ & 1.6 & $3.0 \times 10^{-13}$ & 2.5 & $1.0 \times 10^{-13}$ & 0.9 & $ 2.2 \times 10^{-13}$ & 1.8 & 3.6 & 3.7 \\ \midrule
Bd &  $2.9 \times 10^{-13}$ & 2.4 & $3.0 \times 10^{-13}$ & 2.5 & $1.0 \times 10^{-13}$ & 0.9 & $ 2.2 \times 10^{-13}$ & 1.8 & 4.0 & 4.4 \\ \midrule
Ur &  $3.6 \times 10^{-13}$ & 3.0 & $3.0 \times 10^{-13}$ & 2.5 & $1.0 \times 10^{-13}$ & 0.9 & $ 2.2 \times 10^{-13}$ & 1.8 & 4.4 & 5.0 \\ \midrule
Sh &  $5.9 \times 10^{-13}$ & 4.9 & $3.0 \times 10^{-13}$ & 2.5 & $1.0 \times 10^{-13}$ & 0.9 & $ 2.2 \times 10^{-13}$ & 1.8 & 5.9 & 6.9 \\ \midrule
Ke &  $8.6 \times 10^{-13}$ & 7.2 & $3.0 \times 10^{-13}$ & 2.5 & $1.0 \times 10^{-13}$ & 0.9 & $ 2.2 \times 10^{-13}$ & 1.8 & 7.9 & 8.8 \\ \midrule
Ks &  $1.2 \times 10^{-12}$ & 10.4 & $3.0 \times 10^{-13}$ & 2.5 & $1.0 \times 10^{-13}$ & 0.9 & $ 2.2 \times 10^{-13}$ & 1.8 & 10.9 & 11.8 \\
\bottomrule
\end{tabular}
\label{tab:thermalNoise}
\end{table*}

Once the frequency residual uncertainties have been calculated we proceed to propagate them through the processing pipeline. We assume that the sampled $f_{sky}$ are independent and uncorrelated, hence the frequency covariance matrix $C_f = \langle \Delta f \Delta f^T \rangle$ is diagonal. In order to derive the uncertainties of the ray path parameters, the equations that describe the occultation geometry (Eqs. \ref{eq:a} and \ref{eq:alpha}) are linearized with respect to the ray path angles $\delta_r$ and $\beta_r$ as described by \citet{Fjeldbo1971}, and using the standard propagation of errors \citep{Brandt1997}, the covariance matrix between the bending angle $\alpha$ and the impact parameter $a$, $C_{\alpha a}$, can be derived as follows,

\begin{equation}
C_{\alpha a} = M_{\alpha f} C_f M_{a f}^T
\label{eq:covalphaa}
\end{equation}
where, 
\begin{equation*}
M_{\alpha f} = \left[ \!\begin{array}{c c c c} 
            \frac{\partial \alpha_1}{ \partial f_1 } & \frac{\partial \alpha_1}{ \partial f_2} & \cdots &
\frac{\partial \alpha_1}{ \partial f_n} \\            
            \frac{\partial \alpha_2}{ \partial f_1 } & \frac{\partial \alpha_2}{ \partial f_2} & \cdots &
\frac{\partial \alpha_2}{ \partial f_n} \\ 
\vdots & \vdots & \ddots & \vdots \\
            \frac{\partial \alpha_n}{ \partial f_1 } & \frac{\partial \alpha_n}{ \partial f_2} & \cdots &
\frac{\partial \alpha_n}{ \partial f_n} \\ 
            \end{array} \\
    \!\right]_{f = \bar{f}}
\end{equation*}
and,

\begin{equation*}
M_{a f} = \left[ \!\begin{array}{c c c c} 
            \frac{\partial a_1}{ \partial f_1 } & \frac{\partial a_1}{ \partial f_2} & \cdots &
\frac{\partial a_1}{ \partial f_n} \\            
            \frac{\partial a_2}{ \partial f_1 } & \frac{\partial a_2}{ \partial f_2} & \cdots &
\frac{\partial a_2}{ \partial f_n} \\ 
\vdots & \vdots & \ddots & \vdots \\
            \frac{\partial a_n}{ \partial f_1 } & \frac{\partial a_n}{ \partial f_2} & \cdots &
\frac{\partial a_n}{ \partial f_n} \\ 
            \end{array} \\
    \!\right]_{f = \bar{f}}
\end{equation*}
where $n$ is the number of sampled sky frequencies.
    
The uncertainties associated with the refractive index $\mu$ are determined as explained in \citep{Lipa1979}, using the covariance matrix $C_{\alpha a}$. First, Eq. (\ref{eq:muR0k}) is solved using the trapezoidal approximation and the embedded exponential function is linearized about zero. Then the result is linearized with respect to $\alpha$ and $a$ yielding,

\begin{equation}
\Delta \mu_i = \sum\limits_{k=i}^K (M_{\mu \alpha,k} \Delta \alpha + M_{\mu a,k} \Delta a)
\end{equation}
for the $i$-th concentric spherical layer, where $M_{\mu \alpha, k} =  h_{k-1} - h_k$ and $M_{\mu a, k} = (\partial h_k/\partial a_k)(a_{k+1} - \alpha_k) + (\partial h_{k-1}/ \partial a_k)(\alpha_k - \alpha_{k-1})$ are the linear transformations (to the first order) relating $\mu$ and $\alpha$, and, $\mu$ and $a$, respectively, where $h = \ln{((a_{k+1} + a_k)/2a_i)}$.

The covariance matrix of $\mu$, $C_{\mu}$, is given by,

\begin{align}
C_{\mu} & = \langle \Delta \mu_k \Delta \mu_j \rangle \nonumber \\
        & = \sum\limits_{i=1}^K [V_{\alpha_i}M_{\mu \alpha,ji}M_{\mu \alpha,ki} + V_{a}M_{\mu a,ji}M_{\mu a,ki}  \nonumber \\
        & + C_{\alpha a, i}(M_{\mu \alpha,ki}M_{\mu a,ji} + M_{\mu \alpha,ji}M_{\mu a,ki})]
\label{eq:covmu}
\end{align}
where $V_{\alpha}$ and $V_a$ are the variances of $\alpha$ and $a$, respectively.

From Eq. (\ref{eq:tempN}) two types of uncertainty can be identified for the temperature profile. The first term from the right side of Eq. (\ref{eq:tempN}) results in a systematic error due to the uncertainty of the boundary temperature. The uncertainty resulting from the second term is due to the statistical fluctuation of the refractivity. Using the linearized transformations of Eq. (\ref{eq:tempN}) and Eq. (\ref{eq:idealgas}), the covariance matrices for temperature $C_{T}$ and pressure $C_P$ are derived using Eq. (\ref{eq:covmu}) as described by \citet{Lipa1979} (Appendix C),

\begin{align}
C_T & = \langle \Delta T_k \Delta T_j \rangle \nonumber \\
    & = \sum \limits_{r = k}^{B} \sum \limits_{s = j}^{B} b_r b_s \left( \frac{N_r N_s}{N_k N_j} \right) \left(\frac{C_{\mu,rs}}{N_r N_s} - \frac{C_{\mu,ks}}{N_k N_s} - \frac{C_{\mu,rj}}{N_r N_j} + \frac{C_{\mu,kj}}{N_k N_j} \right)
\label{eq:covt}
\end{align}

\begin{equation}
C_P = \langle \Delta P_k \Delta P_j \rangle = k N_B \sum \limits_{r=k}^B \sum \limits_{s=j}^B b_r b_s C_{\mu,rs}
\label{eq:covp}
\end{equation}
where $b = g(h_{i+1} - h_i)/k$, $g$ is the gravitational acceleration of the central body, which is assumed to be constant, $k$ is Boltzmann's constant and $B$ corresponds to the layer centered at $h_0$ given by the upper boundary condition in Eq. (\ref{eq:tempN}). 

\begin{figure*}[!htp]
    \centering
    \includegraphics[width=\textwidth]{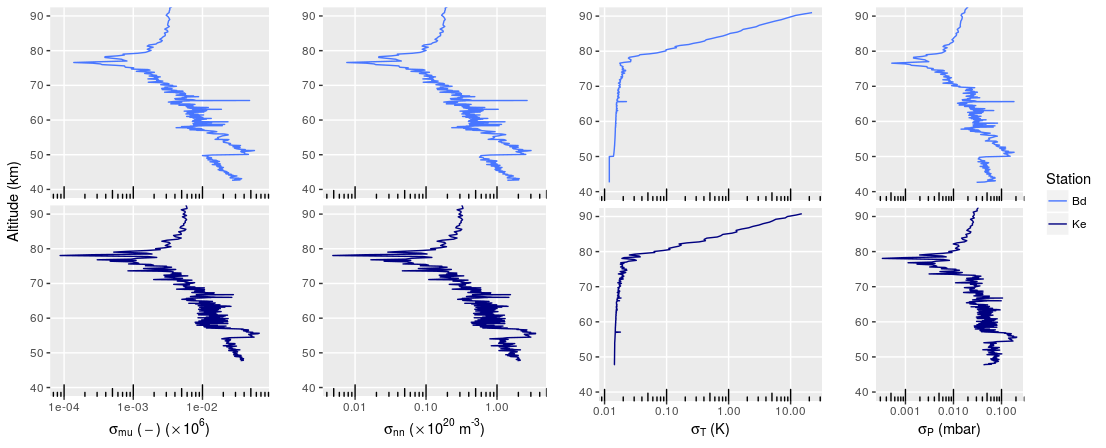}
    \caption[Uncertainties in the refractivity, neutral number density, temperature and pressure profiles, from the error propagation of the Doppler frequency residuals for the stations Bd and Ke during the session of 2012.04.29.]{Uncertainties in the refractivity, neutral number density, temperature and pressure profiles, from the error propagation of the Doppler frequency residuals for the stations Bd and Ke during the session of 2012.04.29. These uncertainties correspond to the profiles shown in Figure \ref{fig:allPlots120429}. The sudden drop at 50 \,km and 55\,km, for Bd and Ke, respectively, are related to the change of integration time from 0.1\,s to 1.0\,s during the processing. }
    \label{fig:sdall}
\end{figure*}

Figure \ref{fig:sdall} shows the resulting uncertainties in the refractivity, neutral number density, temperature and pressure profiles, from the error propagation of the Doppler frequency residuals for the stations Bd and Ke during the session of 2012.04.29. These uncertainties correspond to the profiles shown in Figure \ref{fig:allPlots120429}. 
The minimum values for $\sigma_{\mu}$, $\sigma_{nn}$ and $\sigma_{P}$ are found at an altitude of $\sim$78\,km, which corresponds to the altitude where the temperature profiles of the different stations converge in Figure \ref{fig:sdall} (where the temperature difference between stations drops below 1\,K). The refractivity and neutral number density uncertainties increase approximately linearly from 78\,km to 67\,km. From this altitude to about 63\,km sharp changes are observed at the same altitudes for both stations, which correspond to the altitudes where there are also large changes in the lapse rate as shown in the temperature profile of Figure \ref{fig:allPlots120429}. At $\sim$58\,km, where the tropopause is expected, the refractivity and neutral number density uncertainties start increase at a larger rate as the altitudes decreases to 50 \,km in the case of Bd, and 55\,km in the case of Ke.  The rate at which the uncertainties increase is higher for Ke than for Bd. The sudden drop at 50 \,km and 55\,km, for Bd and Ke, respectively, are related to the change of integration time from 0.1\,s to 1.0\,s during the processing.  The temperature uncertainties are 15\,K and 23\,K, for Bd and Ke, respectively, at 90\,km and rapidly drop below 1\,K at an altitude of $\sim$85\,km. The observed temperature uncertainties above 85\,km explain the differences observed around $\sim$90\,km in the profiles of Bd and Ke, as shown in Figure \ref{fig:allPlots120429}c. Above 85\,km the main contribution to the resulting temperature uncertainty is given by the systematic error induced by the choice of the boundary temperature at 100\,km. As the refractivity increases, this value gets highly damped below the 85\,km.

It is important to take into account that Figure \ref{fig:sdall} is comparing the residuals obtained with a 32-m antenna with those of a 12-m antenna. Badary at an integration time of 0.1\,s has a frequency residual uncertainty of $\sigma_{\Delta f}=18.4$\,mHz and Katherine of $\sigma_{\Delta f}=36.8$\,mHz. The largest uncertainties in the neutral number density found were $\sigma_{nn}=3.1\times10^{20}$\,m$^{-3}$ at 51.2\,km for Bd and $\sigma_{nn}=3.5\times10^{20}$\,m$^{-3}$ at 55.7\,km for Ke. While the ratio of the frequency uncertainties between these two stations is 2, the ratio of the largest neutral number density uncertainties is 1.1. \citet{Tellmann2009} reported neutral density uncertainties $\sigma_{nn}=3.0\times10^{20}\,$m$^{-3}$ at 50\,km with NNO.

\section{Conclusions}
\label{sec:conclusions}

With the VEX test case exposed in this work and the corresponding error analysis, we have demonstrated that the PRIDE setup and processing pipeline is suited for radio occultation experiments of planetary bodies. The noise budget indicated that the uncertainties in the derived neutral density and temperature profiles remain within the range of uncertainties reported in previous Venus radio occultation experiments (\emph{e.g.}, \citet{Tellmann2009}). Summing up the results of all the observations with the different telescopes, we found that at 1 bar level the frequency residuals vary between 4800-5000\,Hz with uncertainties of 3.7-11.8\,mHz, which result in uncertainties in the neutral number densities of 2.7-3.9$\times 10^{20}$\,m$^{-3}$ and in temperature of $\sim$0.01\,K. When characterizing the different sources of Doppler noise, it was found that for one-way radio occultation experiments the noise introduced by the USO can dominate over the thermal noise of large dish antennas (>35m). In the case of the observations with the 65\,m Tianma, this corresponds to 45\,\% of the noise budget. This could be mitigated by performing two-way radio occultation experiment or by using a USO with higher frequency stability such as the Deep Space Atomic Clock (DSAC) \citep{Tjoelker2016}.

Radio occultation experiments carried out with PRIDE can exploit the advantage of having access to large radio telescopes from the global VLBI networks, such as the 65-m Tianma, 100-m Effelsberg or the 305-m Arecibo. Additionally, due to the wide coverage of the networks, the setup can be optimized to ensure high S/N signal detections. For instance, by choosing as receiving stations VLBI telescopes that can track the spacecraft at the highest antenna elevations, when the deep space station has limitations in terms of antenna elevation.

As demonstrated with the detection of Bd, open-loop Doppler data as the one produced with PRIDE allows sounding deeper layers of planetary bodies with thick atmospheres when compared to closed-loop Doppler data. The main advantage of open loop data for radio occultation experiments is that during the post-processing the frequency of the carrier signal can be estimated with precision wideband spectral analysis. Even if there are large unexpected changes in the carrier frequency due to, for instance, large refractivity gradients in the deep atmosphere or interference effects such as multipath propagation. This is not the case with closed-loop detections, since in this scheme the signal is received at much narrower bandwidth. Using a feedback loop, the detection bandwidth is gradually shifted around a central frequency, that is the predicted Doppler signal for the experiment. In case of large unexpected changes in frequency, the signal will no longer be detected by the tracking station because of a loss-of-lock in the closed-loop scheme. With the wideband spectral analysis of PRIDE, we showed that even with small antennas, such as the 12-m Ke, the signal can be detected below Venus' clouds layer.



\begin{acknowledgements}
We thank the referee for her/his constructive comments and corrections of our manuscript, which resulted in an overall improvement of the paper.
The European VLBI Network is a joint facility of independent European, African, Asian, and North American radio astronomy institutes. Scientific results from data presented in this publication are derived from the following EVN project codes: v0427, v0429, v0430, v0501 and v0323. This study made use of data collected through the AuScope initiative. AuScope Ltd is funded under the National Collaborative Research Infrastructure Strategy (NCRIS), an Australian Commonwealth Government Programme. Venus Express (VEX) was a mission of the European Space Agency. The VEX a priori orbit, Estrack and DSN tracking stations transmission frequencies, and the events' schedules were supplied by the ESA’s Venus Express project. The authors would like to thank the personnel of all the participating radio observatories. In particular, the authors are grateful to Eiji Kawai and Shingo Hasagawa for their support of observations at the Kashima radio telescope. The authors are grateful to the Venus Express Radio Science team, the VeRa PI Bernd H\"{a}usler and Venus Express Project Scientists Dmitri Titov and H{\aa}kan Svedhem for their efforts, advice and cooperation in conducting the study presented here. Tatiana Bocanegra-Baham\'{o}n acknowledges the NWO–ShAO agreement on collaboration in VLBI (No. 614.011.501). Giuseppe Cim\`{o} acknowledges the EC FP7 project ESPaCE (grant agreement 263466). Lang Cui thanks for the grants support by the program of the Light in China’s Western Region (No. YBXM-2014-02), the National Natural Science Foundation of China (No. 11503072, 11573057,11703070) and the Youth Innovation Promotion Association of the Chinese Academy of Sciences (CAS).

\end{acknowledgements}

%
%

\bibliographystyle{References/aa}
\bibliography{References/paperVEX.bbl}

\end{document}